# GPU-Accelerated Monte Carlo Simulation and Experimental Study of Radiative Transfer in Multiple Scattering Media

Binhan Wang[a], Peng Sun[a,b,*], Gao Wang[a,b,*], Haijian Liang[b], Jinge Guan[a], Xiaohang Dong[a], Xiuhao Du[a], Ruichen Liu[a]

[a] School of Information and Communication Engineering, North University of China
[b] State key Laboratory of Extreme Environment Optoelectronic Dynamic Measurement Technology and Instrument

**Abstract**

Addressing the problem of photon multiple scattering interference caused by turbid media in optical measurements, biomedical imaging, environmental monitoring and other fields, existing Monte Carlo light scattering simulations widely adopt the Henyey-Greenstein (H-G) phase function approximation model. However, traditional computational resource limitations and high numerical complexity have constrained the application of precise scattering models. Moreover, the single-parameter anisotropy factor assumption neglects higher-order scattering effects and backscattering intensity, failing to accurately characterize the multi-order scattering properties of complex media. To address these issues, we propose a GPU-accelerated Monte Carlo-Rigorous Mie scattering transport model for complex scattering environments. The model employs rigorous Mie scattering theory to replace the H-G approximation, achieving efficient parallel processing of phase function sampling and complex scattering processes through pre-computed cumulative distribution function optimization and deep integration with CUDA parallel architecture. To validate the model accuracy, a standard scattering experimental platform based on 5μm polystyrene microspheres was established, with multiple optical depth experimental conditions designed, and spatial registration techniques employed to achieve precise alignment between simulation and experimental images. The research results quantitatively demonstrate the systematic accuracy advantages of rigorous Mie scattering phase functions over H-G approximation in simulating lateral scattering light intensity distributions, providing reliable theoretical foundations and technical support for high-precision optical applications in complex scattering environments.

parallel computing; turbid media; radiative transfer

# 1. Introduction

## 1.1 Research Background and Significance

With the rapid development of modern optical technology, research on light transport characteristics in complex scattering media has become increasingly important. In numerous fields including biomedical optical imaging, environmental optical remote sensing, industrial process monitoring, and marine optical research, the presence of turbid media significantly affects light transmission paths and intensity distributions, directly impacting system signal quality and imaging quality.

In practical applications, experiments are often conducted in complex environments containing turbid media such as suspended particles, aerosols, fog droplets, and biological tissues[1,2]. These media significantly interfere with light signal transmission paths through scattering, absorption, and attenuation effects, leading to problems such as reduced imaging contrast and decreased signal-to-noise ratio. In biomedical imaging, the strong scattering properties of biological tissues limit the penetration depth and resolution of optical imaging; in environmental monitoring, atmospheric aerosols and haze affect the detection accuracy of laser radar; in industrial inspection, suspended particles interfere with the measurement results of optical sensors; in marine optics, scattering particles in water bodies affect the effectiveness of underwater imaging.

Monte Carlo methods are widely applied in photon transport simulation. Berrocal et al.[3,4] established a light scattering transport model based on Monte Carlo methods, laying the theoretical foundation for laser scattering research in turbulent media, but simulation efficiency was low due to limitations in CPU computational capability. A. Collin et al.[5] established a numerical model for water mist radiation attenuation, employing Monte Carlo methods to handle radiative transfer, but due to computational resource constraints, the grid resolution was relatively low, making it difficult to meet the demands of large-scale refined simulation. Erik Alerstam et al.[6] implemented a GPU-accelerated version of MCML, primarily applicable to layered homogeneous media models, using macroscopic scattering coefficients to describe light transport, which cannot precisely handle the Mie scattering characteristics of discrete particles. Young-Schultz[7] achieved significant GPU acceleration in biological tissue light transport using FullMonteCUDA, but it employs H-G phase function approximation for scattering processes, unable to precisely describe the complex angular distribution characteristics of Mie scattering, making it difficult to meet the requirements of high-precision optical applications. The Multi-Scattering online platform developed by Jönsson and Frantz[8,9] employs GPU cluster acceleration, achieving significant computational speedup. However, its online version is currently limited to 20×20×20 voxel grids, mainly constrained by the complexity of data input through the web interface, which to some extent limits the refined modeling capability for complex geometric structures. Zhang[10] established a Monte Carlo model based on dynamic particle parameter selection for simulating light transport in fog media, where the model dynamically selects optical parameters of medium particles through collision probability. Although this method achieved improvements in simulation accuracy and computational efficiency, it was primarily validated through numerical simulation, lacking sufficient experimental validation to quantitatively assess the model's accuracy in actual fog environments.

Due to early Monte Carlo photon transport simulations being constrained by computational resources, compromises could only be made between real-time performance and precision, leading most studies to adopt H-G phase function approximations. Existing H-G phase function models, due to their single-parameter anisotropy factor assumption of axially symmetric scattering phase distribution, ignore higher-order Legendre terms and cannot characterize multi-order scattering effects in complex media, neglecting backscattering intensity. Additionally, transient medium concentration variations make it difficult for static anisotropy factor values to respond in real-time, resulting in correction lag. Although this function is computationally simple and easy to implement, it fails to capture the typical oscillatory structures and complex angular distributions characteristic of Mie scattering. To compensate for errors introduced by this physical simplification, we introduce rigorous Mie scattering theory into a GPU-accelerated Monte Carlo framework, utilizing analysis of simulation images from different phase functions against experimental images to demonstrate the necessity and feasibility of rigorous Mie scattering in refined modeling of complex media. This approach not only overcomes the physical limitations of H-G approximation but also fully meets the demands of high-precision optical measurement and imaging applications.

To validate simulation accuracy, we established a standard scattering experimental platform based on 5μm polystyrene microspheres, designing multiple experimental conditions with optical depths (OD) ranging from 5 to 12.5. Due to factors such as geometric alignment errors in experimental apparatus, camera positioning deviations, and coordinate system differences between simulation and experiment, GPU-accelerated spatial registration techniques were employed to perform translational transformations on images, eliminating spatial position deviations and achieving precise alignment between simulation and experimental images. Through multidimensional quantitative metrics including root mean square error and radial error analysis, the systematic accuracy advantages of rigorous Mie scattering phase functions over H-G approximation in simulating lateral scattering light intensity distributions were quantitatively validated.

## 2. Theoretical Foundations

### 2.1 Radiative Transfer Theory

In scattering processes, optical depth (OD) is introduced as a parameter to describe the total attenuation capability of a medium for light, incorporating both scattering and absorption effects. Three scattering regimes can be identified based on optical depth[4]. Single scattering regime corresponds to OD ≤ 1, where single scattering events dominate. Intermediate scattering regime applies to 2 ≤ OD ≤ 9, where visibility decreases and the dominant number of scattering events (also called scattering order) typically approaches the OD value. For OD ≥ 10, this belongs to the multiple scattering regime, where standard imaging methods lose visibility[9]. When incident light is parallel monochromatic light perpendicularly illuminating a uniform medium containing scattering and absorption centers, according to the Beer-Lambert law, light undergoes exponential attenuation with propagation distance $l$ and extinction coefficient $\mu_{ext}$ in the medium, deriving the transmitted light intensity $I$[8] as shown in equation (1)

$$I = I_0 exp^{-OD} \text{ and } OD = \mu_{ext} l \tag{1}$$

where $\mu_{ext} = \mu_{sca} + \mu_{abs}$, $\mu_{sca}$ and $\mu_{abs}$ are the scattering and absorption coefficients,

respectively.

The Beer-Lambert law applies to non-scattering or weakly scattering media. For turbid media where significant absorption and scattering coexist, the radiative transfer equation is needed to describe the spatial and directional distribution of light intensity[8]. Radiative transfer theory is based on energy conservation of incident, absorbed, scattered, and emergent photons within infinitesimal volume elements. The radiative transfer equation can be described as follows: the change in radiance along the incident direction corresponds to the radiance lost due to extinction of incident light, plus the total amount of radiance scattered from all other directions into the incident direction. This is an extended form of the Beer-Lambert law that considers the redistribution effects of scattered light, as shown in equation (2).

$$\partial I(\boldsymbol{r},\boldsymbol{s})/\partial s = -\mu_{ext}I + (\mu_{sca}/4\pi)\int_{4\pi} P(\boldsymbol{s},\boldsymbol{s}')I(\boldsymbol{r},\boldsymbol{s}')d\Omega' \qquad (2)$$

where $\boldsymbol{r}$ is position, $\boldsymbol{s}$ is direction, and $d\Omega'$ is the solid angle differential element. In equation (2), $-\mu_{ext}I$ represents the light intensity loss along the incident light direction equal to the intensity loss caused by scattering and absorption of incident light in other directions, while the integral term represents the total amount of intensity increase caused by scattering back from other directions to the original incident direction.

**2.2 Mie scattering Theory**

Mie scattering theory describes the precise physical process of light wave interaction with spherical particles and is the core theory for handling scattering problems where particle size is close to or larger than the incident light wavelength. The essence of the scattering process can be viewed as elastic collision between radiation photons and scattering elements. Scattering laws vary with different sizes of scattering bodies. In scattering, a size parameter $x$[11] is commonly used as shown in the equation:

$$x = \pi d/\lambda \qquad (3)$$

where $d$ is the diameter of the scattering particle and $\lambda$ is the incident light wavelength. When $x \geq 0.1$, scattering follows Mie scattering laws, which is the precise theory describing light wave scattering by spherical particles, applicable to particles with sizes close to or greater than one-tenth of the incident light wavelength, such as aerosols, water droplets, smoke particles, etc. Mie scattering theory has no strict theoretical upper limit, and when $x \gg 1$, it exhibits strong forward scattering characteristics[12].

The core of Mie scattering theory lies in solving Maxwell's equations, where single particle scattering, scattering phase function, and anisotropy factor are key parameters in Mie scattering theory. Single particle scattering concerns the scattering behavior of individual particles to incident light, and Mie scattering theory is the core theory describing this process. Single particle scattering involves two key parameters: the complex refractive index $m$ and the size parameter $x$. The complex refractive index $m$[10] is given by equation (4)

$$m = n + ik \tag{4}$$

where $n$ is the real part of the refractive index and $k$ is the imaginary part of the complex refractive index. The real part $n$ determines the propagation speed of light waves in absorbing media, while the imaginary part $k$ determines the attenuation of light waves when propagating in absorbing media.

The core calculation of Mie scattering involves two scattering amplitude functions: the scattering amplitude function for perpendicular polarization $S_1(\theta)$ as shown in equation (5) and the scattering amplitude function for parallel polarization $S_2(\theta)$[13] as shown in equation (6).

$$S_1(\theta) = \sum_{n=1}^{\infty} (2n+1)/(n(n+1)) [a_n \pi_n(\cos \theta) + b_n \tau_n(\cos \theta)] \tag{5}$$

$$S_2(\theta) = \sum_{n=1}^{\infty} (2n+1)/(n(n+1)) [a_n \tau_n(\cos \theta) + b_n \pi_n(\cos \theta)] \tag{6}$$

where $a_n$ and $b_n$ are Mie scattering coefficients, and $\pi_n$ and $\tau_n$ are angular functions related to Legendre polynomials. The calculation of Mie coefficients[14] is shown in the following equations:

$$a_n = (m\psi_n(mx)\psi_n'(x) - \psi_n(x)\psi_n'(mx))/(m\psi_n(mx)\xi_n'(x) - \xi_n(x)\psi_n'(mx)) \tag{7}$$

$$b_n = (\psi_n(mx)\psi_n'(x) - m\psi_n(x)\psi_n'(mx))/(\psi_n(mx)\xi_n'(x) - m\xi_n(x)\psi_n'(mx)) \tag{8}$$

From the Mie amplitude functions $S_1(\theta)$ and $S_2(\theta)$, the scattering phase function $P(\cos \theta)$[15] can be derived, which can describe the distribution probability and scattering capability of photons along different directions, calculated as equation (9)

$$P(\cos \theta) = (1/k^2)(|S_1(\theta)|^2 + |S_2(\theta)|^2) \tag{9}$$

Where $k = 2\pi m_{med}/\lambda$ is the wave number in the medium. To ensure probability conservation, according to literature[16] normalization processing is needed in $P(cos\ \theta)$ as shown in equation (10)

$$1/(4\pi) \int_{4\pi} P(cos\ \theta) d\Omega = 1 \qquad (10)$$

where $\theta$ is the scattering angle and $d\Omega = sin\theta d\theta d\phi$. If normalization processing is not performed, simulation results will deviate from real physical laws.

The Mie scattering phase function exhibits complex angular dependence, with its characteristics showing clear evolutionary patterns as the size parameter $x$ varies. In the Rayleigh scattering region ($x \ll 1$), the phase function approaches isotropic distribution; as $x$ gradually increases entering the transition region ($x \approx 1$), forward scattering gradually strengthens; when entering the Mie scattering region ($x \gg 1$), forward scattering significantly strengthens, backscattering obviously weakens, and complex oscillatory structures appear in the medium-to-large angle range.

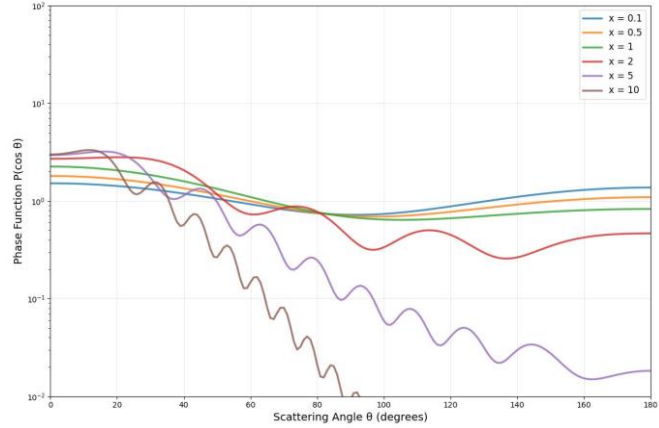

Fig. 1. The variation of Mie scattering phase function with $x$

### 2.3 Henyey-Greenstein Phase Function

The H-G phase function is an empirical approximation model widely applied in radiative transfer simulations. This function is defined as:

$$f(\theta) = (1/(4\pi)) \left((1 - g^2)/(1 + g^2 - 2g \cdot \cos(\theta))^{3/2}\right) \qquad (11)$$

where $g$ is the anisotropy factor, representing the average cosine value of the scattering angle. The H-G function depends only on a single parameter $g$. Although it is computationally simple and easy to implement, the H-G function cannot accurately describe the complex angular distribution characteristics of rigorous Mie scattering. In applications such as high-precision optical measurements and biomedical imaging, its inherent theoretical limitations seriously affect the accuracy of radiative transfer simulations, particularly in the oscillatory structures within medium-to-large angle ranges. In contrast, rigorous Mie scattering theory can accurately capture detailed features at all scattering angles, providing higher modeling precision for precise optical measurements and imaging analysis[8].

### 2.4 Anisotropy Factor

The anisotropy factor $g$ is an important characteristic parameter of the scattering phase function, defined[17] as:

$$g = (1/2) \int_{-1}^{1} P(\cos\theta)\cos\theta \, d(\cos\theta) \tag{12}$$

The value range is $-1 \leq g \leq 1$. When $g = 1$, light propagates completely along the incident direction; when $g = -1$, light scatters completely in the backward direction; when $g = 0$, isotropic scattering occurs. When Mie scattering occurs, the value of $g$ is close to 1. In Mie scattering, the value of $g$ depends on the particle size parameter $x$ and directly affects the efficiency of Monte Carlo simulations.

### 2.5 Extinction Cross-Section Calculation

The extinction cross-section $\sigma_e$ is a key parameter for controlling OD[18], For 5μm monodisperse polystyrene microspheres, rigorous calculation through Mie scattering theory was performed. According to literature[19] the refractive index is $m = 1.594 - 0.00033i$, and the refractive index of surrounding water is $m = 1.333 - 0.0i$. The size parameter $x$ employs the Mie coefficient recurrence algorithm proposed by Bohren and Huffman[12] Through the extinction efficiency factor formula:

$$Q_{ext} = (2/x^2) \sum_{n=1}^{\infty} (2n+1)\text{Re}(a_n + b_n) \tag{13}$$

$Q_{ext}= 2.25$ was obtained, and the theoretical extinction cross-section was further calculated:

$$\sigma_e = Q_{ext} \cdot \frac{\pi d^2}{4} = 2.25 \times \frac{\pi (5\mu m)^2}{4} = 4.42 \times 10^{-5} \text{mm}^2 \tag{14}$$

### 2.6 Monte Carlo Radiative Transfer Model

The Monte Carlo method is a statistical simulation technique that is well-suited for handling complex physical problems. In research involving tracking large numbers of photons in turbid media transport[8], the Monte Carlo method treats each photon as an independent particle, with its transport process described through a series of random events including free propagation, scattering, absorption effects, and boundary processing. The complexity of turbid media structure directly affects the complexity of the Monte Carlo model required for simulation. Turbid media are characterized by particle concentration, their spatial distribution, and the number of different particle sizes or types present[20].By incorporating physical parameters from Mie scattering theory, efficient modeling of photons for multiple scattering effects can be achieved. This method reproduces the overall radiative transfer phenomena through statistical tracking of large numbers of individual photons.

When photons undergo scattering interaction after propagating distance $s$, the scattering angle $\theta$ is sampled according to the Mie scattering phase function $P(\cos\theta)$, and the azimuthal angle $\phi$ is uniformly randomly sampled from $[0,2\pi)$. Pre-computed cumulative distribution function (CDF)[21] inverse transform is employed to generate the scattering angle $\theta$ to improve computational efficiency, converting the traditional method that requires real-time solving of integral equations into efficient look-up table operations.

### 2.7 GPU Parallelization Implementation

In the simulation process, to obtain convergent results with hundreds of millions of photons in simulation, we employ GPU-parallelized Monte Carlo algorithms, implementing parallel processing of photon quantities through CUDA technology, with each CUDA thread independently handling the complete transport process of one photon. Through optimized parallel strategies, the GPU version achieves computational speed improvements of hundreds of times, enabling rapid accurate simulation of hundreds of millions or more photons.

## 3. Numerical Simulation Methods

### 3.1 Algorithm Implementation

We utilize a GPU-accelerated Monte Carlo-Rigorous Mie scattering simulation algorithm to simulate radiative transfer characteristics. The algorithm simulation is divided into four modules: parameter processing module, rigorous Mie scattering calculation module, Monte Carlo calculation module, and data analysis and visualization module. The complete Monte Carlo simulation process can be referenced in Fig. 2.

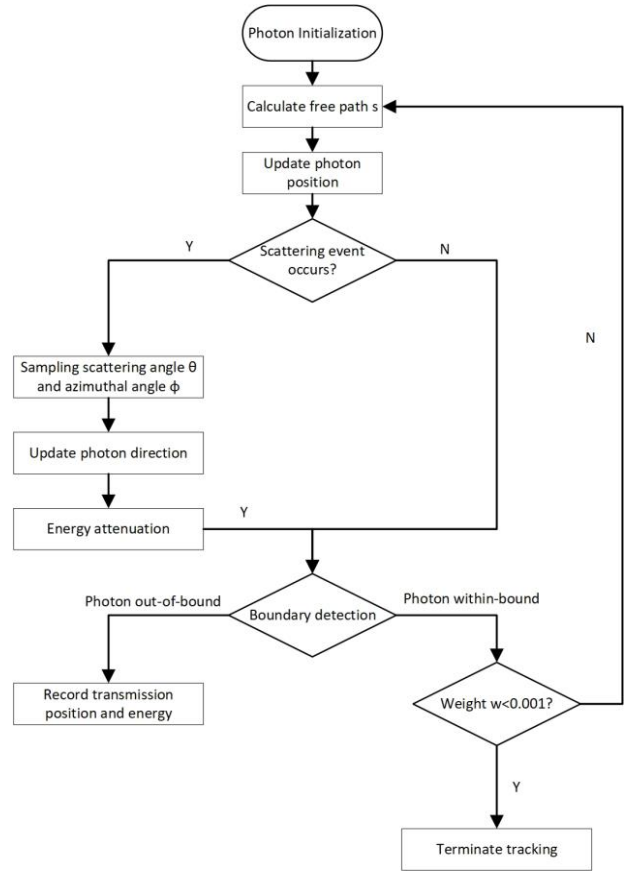

Fig. 2. Monte Carlo simulation flowchart

### 3.1.1 Parameter Processing Module

The parameter processing module is responsible for receiving and preprocessing various physical parameters required for simulation, including input particle size diameter $d$, wavelength $\lambda$, and OD, calculating size parameter $x$, scattering coefficient $\mu_{sca}$, and absorption coefficient $\mu_{abs}$. It also has parameter validity verification functionality to ensure that input parameters are within reasonable ranges, providing accurate and reliable input for subsequent simulation calculations.

### 3.1.2 Rigorous Mie scattering Calculation Module

The rigorous Mie scattering calculation module is the theoretical core of the entire algorithm, responsible for calculating key physical quantities such as Mie scattering coefficients $a_n$ and $b_n$, scattering phase function $P(\cos\theta)$, and anisotropy factor $g$. To ensure calculation accuracy, we adopt an improved Mie coefficient calculation algorithm. For large size parameter ($x > 10$) cases, traditional recursive algorithms are prone to numerical overflow or underflow problems. We introduce a multi-precision computation library (mpmath) for key calculation steps, improving calculation precision to 50 decimal places, ensuring accuracy and stability of Mie coefficient calculations. Phase function pre-computation and CDF optimization are key technologies for improving simulation efficiency. By pre-computing phase function values at 3000 angular points and constructing cumulative distribution functions, scattering angle sampling is converted from real-time integral calculation to efficient binary search, significantly improving GPU parallel computing efficiency.

### 3.1.3 Monte Carlo Calculation Module

The Monte Carlo calculation module is the execution core of the entire algorithm, responsible for photon initialization, propagation, scattering, and detection processes. This module employs GPU parallel architecture, supporting efficient processing of 4.56 million photons per second. The module design includes the following core components:

Photon Parallel Initialization: Photon state arrays are allocated in GPU global memory, including position, direction, weight, and phase information. Parallel processing of large numbers of photon initializations through CUDA thread blocks achieves efficient memory access patterns.

Adaptive Transport Distance Sampling: In simulation, photons are assumed to be perpendicularly incident from the center of the medium along the x-axis direction. For parallel light beams, all photons have the same initial direction. The free propagation distance $s$ of photons is determined by calculating the mean free path length using formula (15)[22]

$$s = -ln(u)/\mu_{ext}, u \sim U(0,1) \tag{15}$$

where the extinction coefficient $\mu_{ext} = \mu_{sca} + \mu_{abs}$. To avoid performance bottlenecks of random number generation in GPU parallel computing, we generate large quantities of random numbers on the CPU side, then transfers them to GPU constant memory for photon parallel sampling use.

Scattering Event Parallel Processing: Each CUDA thread independently processes the complete transport process of one photon, including scattering angle sampling based on pre-computed CDF look-up tables, direction updates, and boundary determination. For comparative analysis, this module simultaneously performs scattering angle sampling according to the standard H-G phase function formula, with its anisotropy factor $g$ set equal to the $g$ value calculated from Mie scattering theory.

Energy Attenuation and Weight Updates: The scattering, absorption, and attenuation processes of photons after each scattering event are achieved through dynamic adjustment of photon weight $w$[23] as shown in equation (16):

$$w_{new} = w_{old} \cdot (1 - (\mu_{abs}/\mu_{ext})) \tag{16}$$

When weight falls below a threshold (such as $w$ < 0.001), photon tracking is terminated to improve computational efficiency. This attenuation method can accurately simulate the absorption effects of media while maintaining energy conservation.

Memory Optimization and Atomic Operations: Through memory access patterns and atomic operations, thread safety is ensured when multiple threads simultaneously update shared data. Particularly in the light intensity distribution statistics process, CUDA atomic addition operations are employed to maintain data consistency.

### 3.1.4 Data Analysis and Visualization Module

The data analysis and visualization module are responsible for post-processing, analysis, and visualization of simulation results. This module employs basic normalization processing methods to ensure that experimental and simulation data are compared and analyzed within the same numerical range. All image data adopt the same normalization processing standards to ensure fair comparison between experimental and simulation data.

### 4 Experimental Methods

The experimental system employs a controlled scattering environment design to ensure precise correspondence between simulation parameters and experimental conditions. The experimental apparatus mainly includes laser light source, scattering medium configuration, optical detection, and data processing.

### 4.1 Laser Light Source

A 532nm continuous laser is employed as the light source, with power stability better than 0.1%. The laser beam forms a parallel beam with a diameter of 2.5mm through a beam expansion and shaping system, with a divergence angle less than 0.5mrad. The laser beam is collimated and expanded to optimize beam quality, with an aperture controlling the beam diameter, ensuring the detector operates within the linear response range.

**4.2 Scattering Medium Configuration**

Due to the fact that in turbid media, droplets easily evaporate or condense, particle size distribution changes dynamically over time, and they are easily affected by airflow disturbances, and when using optical instrument, they easily lead to signal instability, the water mist medium was replaced with monodisperse polystyrene microsphere dispersion with controllable particle size, high monodispersity, and greater stability. The 5μm polystyrene microspheres have high monodispersity, can ensure repeatability of experimental results, and are not easily dissolved in deionized water dispersion, making them more suitable for long-term experiments.

Optical Depth Control: Based on the Beer-Lambert law, precise control of target OD is achieved by diluting polystyrene microsphere stock solution with a solid content of 25g/ml. OD is jointly determined by particle number density $n$, extinction cross-section $\sigma_e$, and optical path length $L$ as shown in equation (17):

$$\text{OD} = n \cdot \sigma_e \cdot L \tag{17}$$

Here $n = C_{\text{diluted}}/(m_p \cdot 10^3)$, $C_{\text{diluted}}$ is the mass concentration after dilution, and $m_p$ is the mass of a single microsphere, calculated from microsphere diameter $d$ and density (ρ = 1.05g/cm³). By solving the dilution equation for stock solution volume ($V_{stock}$, mL) and added water volume ($V_{water}$, mL), formula (18) is derived:

$$V_{water} = V_{stock}\big((C_{\text{diluted}} \cdot \sigma_e \cdot L)/(\text{OD}_{\text{target}} \cdot m_p \cdot 10^3)\big) - V_{stock} \tag{18}$$

The extinction cross-section $\sigma_e = 4.42 \times 10^{-5} mm^2$ (for $d$ =5μm, $\lambda$ =532nm monodisperse polystyrene microspheres) is obtained through theoretical calculation. The model validates its reliability by comparing predicted OD values with measured values under different dilution ratios. In the experiment, 5 groups of experimental conditions with different ODs are designed. After directly calculating dilution ratios through the algorithm, polystyrene microsphere stock solution is diluted to achieve precise control of target optical depth OD from high to low. Graduated pipettes with measuring ranges of 10ml, 5ml, and 1ml (with graduation values of 0.1, 0.05, and 0.01 respectively) are used for volumetric measurement of pure water and polystyrene solution. The solution ratios corresponding to different OD values are shown in Table 1.

Table 1 Solution ratio conditions for different OD values

| OD | 12.5 | 10 | 8 | 7.5 | 5 |
|---|---|---|---|---|---|
| Water/ml | 20.15 | 25.31 | 31.76 | 33.91 | 51.12 |
| Polystyrene/ml | 0.5 | 0.5 | 0.5 | 0.5 | 0.5 |

### 4.3 Optical Detection and Data Processing

A quartz glass cuvette container with dimensions of 30×10×45mm³ and transmittance >99% was selected. Lateral scattering detection employs a 90° geometric configuration, recording lateral scattering light intensity distribution through a high-sensitivity CMOS camera. Grayscale mode shooting is adopted, with exposure time adjusted according to scattering intensity (1-5ms). The experimental setup and apparatus diagram can be referenced in Fig.3.

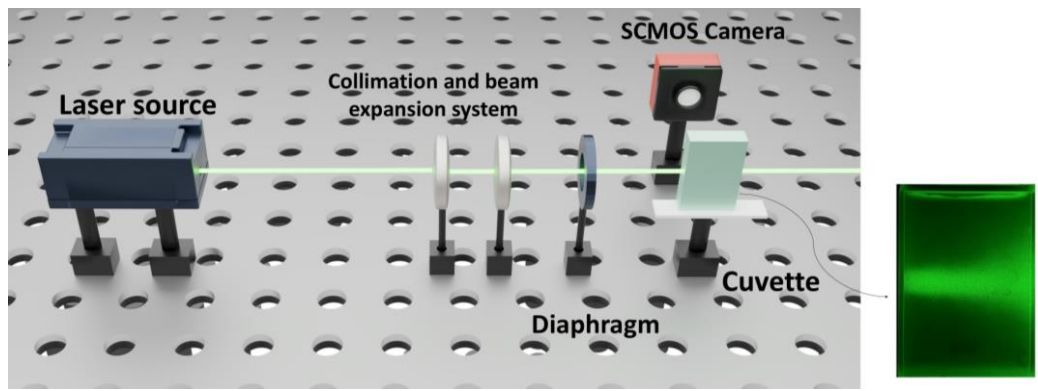

Fig.3. Experimental setup and apparatus diagram

Laser power calibration is performed, with incident light aligned to the center position of the cuvette, and the camera positioned directly facing the side of the cuvette. Detection is conducted sequentially according to OD values from high to low dilution. After normalizing the captured data, correlation coefficients and other data between experimental and simulation images are analyzed.

## 5. Results and Discussion
### 5.1 Theoretical Calculation of Rigorous Mie scattering Parameters and Phase Function Characteristics

Based on the optical parameters of 5μm monodisperse polystyrene microspheres under 532nm laser illumination, key scattering parameters were obtained through calculation. This table presents the key rigorous Mie scattering parameters and their physical meanings for 5μm polystyrene microspheres.

Table 2 Theoretical parameter calculation results

| Parameter | Size Parameter $x$ | Extinction Coefficient $\mu_{ext}$/mm$^{-1}$ | Scattering Coefficient $\mu_{sca}$/mm$^{-1}$ | Absorption Coefficient $\mu_{abs}$/mm$^{-1}$ | Anisotropy Factor $g$ |
|---|---|---|---|---|---|
| Value | 29.5 | 0.268 | 0.267 | 0.001 | 0.8722 |

The size parameter $x$ = 29.5 indicates that the particle size is much larger than the incident wavelength ($x \gg 1$), falling within the Mie scattering regime, where scattering behavior exhibits strong directional characteristics. Through calculation, $\mu_{ext}$ and $\mu_{sca}$ are 0.268 and 0.267 mm$^{-1}$ respectively, with their similar values indicating that light attenuation is primarily caused by scattering. The absorption coefficient is only 0.001 mm$^{-1}$, confirming that polystyrene material has virtually no absorption characteristics at 532nm wavelength. The anisotropy factor $g$ = 0.8722 approaches the theoretical maximum value of 1, indicating that scattered light is mainly concentrated in the forward direction, embodying the typical characteristics of large particle Mie scattering.

Through Rigorous Mie scattering analysis of 5μm polystyrene microspheres, calculation results show significant differences between rigorous Mie scattering phase function and H-G phase function. As shown in Fig. 4., the rigorous Mie scattering phase function presents an extremely sharp peak in the forward scattering direction (0° direction), which forms a striking contrast with the smooth forward scattering peak of the H-G phase function. The rigorous Mie scattering phase function exhibits complex oscillatory structures throughout the entire scattering angle range, particularly showing multiple secondary peaks and fine oscillatory patterns in lateral scattering (90° and 270° directions) and backscattering regions (180° direction). These complex oscillatory features are direct manifestations of fine interference effects between light waves and spherical particles, reflecting the completeness of Mie scattering theory. In contrast, the H-G phase function presents a smooth approximately elliptical distribution that, while capable of reasonably fitting the overall trend of forward scattering, completely neglects the rich oscillatory details in Mie scattering. This difference indicates that for large particle scattering problems with sizes comparable to wavelength, using rigorous Mie theory calculations is crucial, while simplified phase function approximations may lose important scattering characteristic information.

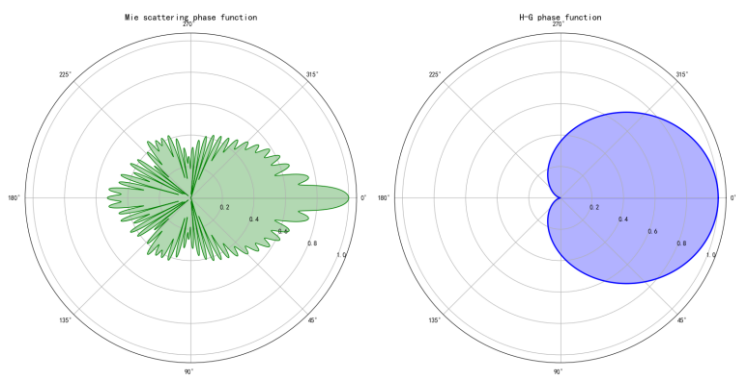

Fig. 4. Polar coordinate scattering diagram of 5μm polystyrene microspheres

**5.2 Simulation Results Validation and Phase Function Performance Comparison**

To systematically evaluate the accuracy of the Monte Carlo simulation model, we conducted qualitative and quantitative comparative analysis of simulation images and experimental images under different OD. Fig. 5. shows rigorous Mie scattering simulation images at different optical depths, with simulation images obtained by simulating $3\times10^8$ photons passing through the cuvette using NVIDIA RTX3050.

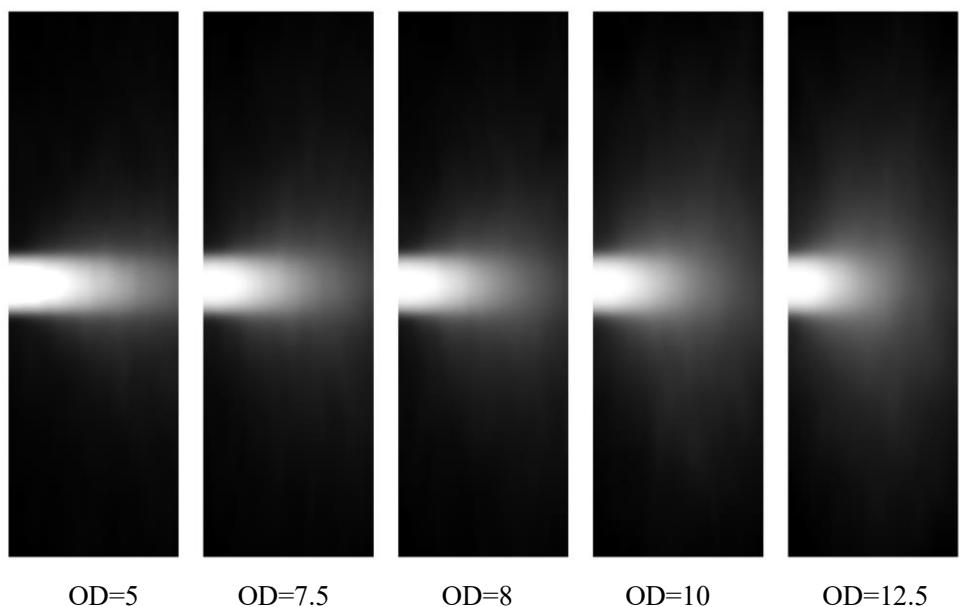

OD=5　　　OD=7.5　　　OD=8　　　OD=10　　　OD=12.5

Fig. 5. Simulation images at different optical depths under rigorous Mie scattering phase function

Fig. 6. comprehensively presents triple image comparisons of experimental measurements, rigorous Mie scattering simulation, and H-G phase function simulation results under different

optical depths. From the image comparison, it can be observed that rigorous Mie scattering simulation images exhibit high consistency with experimental images in both light intensity distribution details and overall morphology. In contrast, while H-G phase function simulation images are similar to experimental results in macroscopic trends, they show obvious deviations in capturing fine structures of light fields and local intensity variations.

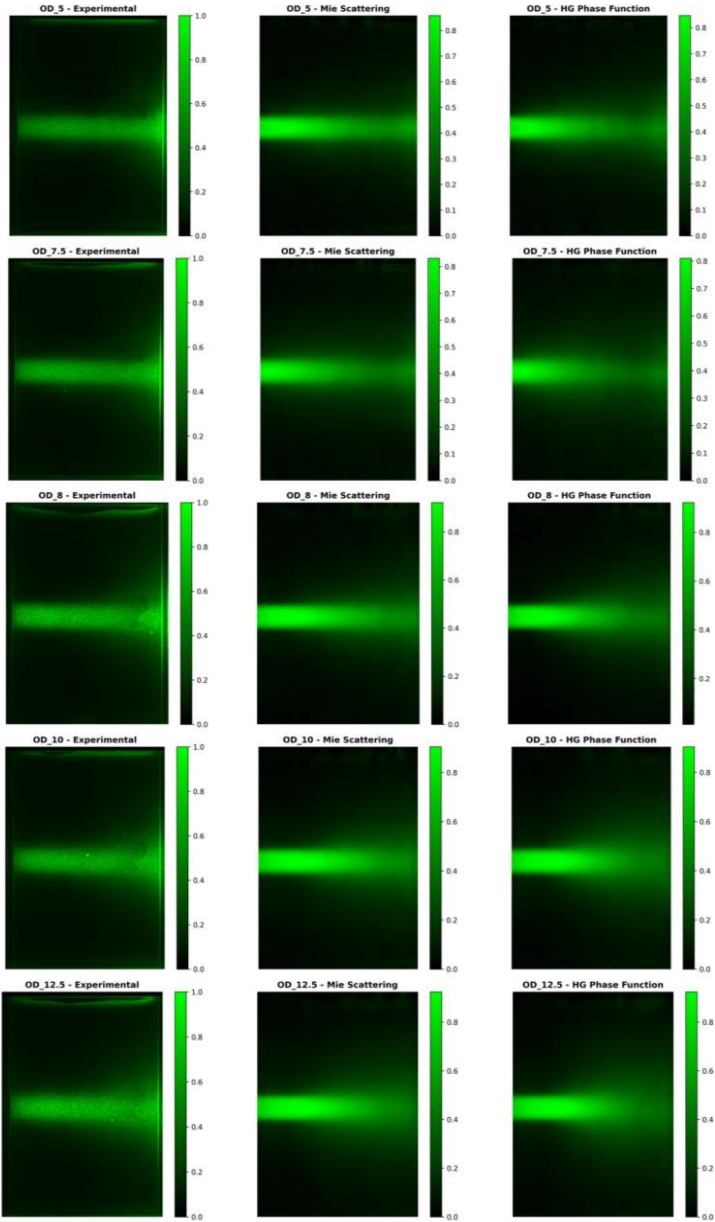

Fig. 6. Post-processed experimental and simulation registered images

Scattering image comparison under different ODs: (left) experimental images (center) rigorous Mie scattering simulation (right) H-G scattering simulation

Table 3 presents quantitative comparative analysis of rigorous Mie scattering models and H-G phase functions using root mean square error (RMSE) and correlation coefficient (r) with experimental data as the benchmark.

Table 3 Comparison of key error metrics between rigorous Mie scattering and H-G phase function at different optical depths

| OD | Metric | H-G | Rigorous Mie scattering | Improvement (%) |
| --- | --- | --- | --- | --- |
| 5 | RMSE | 0.139994 | 0.136938 | 2.18 |
|  | r | 0.6438 | 0.6934 | 7.71 |
| 7.5 | RMSE | 0.136818 | 0.128896 | 5.79 |
|  | r | 0.5942 | 0.6719 | 13.08 |
| 8 | RMSE | 0.136350 | 0.132351 | 2.93 |
|  | r | 0.7539 | 0.7839 | 3.98 |
| 10 | RMSE | 0.137065 | 0.132898 | 3.04 |
|  | r | 0.7601 | 0.7891 | 3.81 |
| 12.5 | RMSE | 0.140049 | 0.134208 | 4.17 |
|  | r | 0.7460 | 0.7813 | 4.73 |
| Average | RMSE | 0.138055 | 0.133058 | 3.62 |
|  | r | 0.6996 | 0.7439 | 6.33 |

The quantitative comparison results in Table 3 confirm the systematic advantages of the rigorous Mie scattering model in simulation accuracy. Under all tested optical depths, rigorous Mie scattering generally exhibits lower RMSE than the H-G phase function and generally higher correlation coefficients. From the average values, the rigorous Mie scattering model achieved an average RMSE improvement of 3.62% and a correlation coefficient improvement of 6.33%.

To deeply evaluate the spatial distribution differences between the two scattering phase functions, we introduce a radial error analysis method, converting two-dimensional scattering images to radial coordinate systems and extracting radial intensity distributions through angular averaging processing. Based on differences in scattering physical mechanisms, the radial region is divided into three characteristic zones: the forward scattering zone (corresponding to small angle scattering θ < 30°), mainly reflecting the strong forward peak characteristics of large particle Mie scattering; the intermediate zone (corresponding to medium angle scattering 30° ≤ θ ≤ 120°), embodying the complex oscillatory structure of Mie scattering; and the edge zone

(corresponding to large angle scattering θ > 120°), reflecting backscattering and edge effects.

Table 4 Comparison of radial error and correlation between two phase function models at different OD

| OD | Forward Scattering Zone | | Intermediate Zone | | Edge Zone | | Radial Correlation | | Mie Better (%) |
|---|---|---|---|---|---|---|---|---|---|
| | Mie | H-G | Mie | H-G | Mie | H-G | Mie | H-G | |
| 5 | 0.0976 | 0.0513 | 0.0813 | 0.0982 | 0.0853 | 0.0899 | 0.9868 | 0.9691 | 79.3% |
| 7.5 | 0.0913 | 0.1520 | 0.0794 | 0.1025 | 0.0920 | 0.1033 | 0.9716 | 0.9360 | 100% |
| 8 | 0.1214 | 0.0944 | 0.0784 | 0.0995 | 0.0770 | 0.0800 | 0.9882 | 0.9784 | 69.0% |
| 10 | 0.1964 | 0.1685 | 0.1019 | 0.1229 | 0.0703 | 0.0795 | 0.9956 | 0.9890 | 79.3% |
| 12.5 | 0.1586 | 0.1353 | 0.1007 | 0.1200 | 0.0765 | 0.0900 | 0.9978 | 0.9934 | 82.8% |

The radial error analysis results indicate that the rigorous Mie scattering model exhibits lower errors under most optical depths, especially in the intermediate and edge regions. Under OD=7.5 conditions, the rigorous Mie scattering shows significantly lower errors than the H-G phase function in the forward scattering zone, intermediate region, and edge region. From the perspective of overall spatial distribution, the rigorous Mie scattering model demonstrates superior performance in intermediate and edge regions as well as overall correlation, while the H-G phase function cannot accurately describe scattering characteristics in lateral and backward scattering regions. This comparison not only validates the physical limitations of the H-G phase function under high OD and complex angular distribution conditions, but also quantitatively proves the necessity and reliability of rigorous Mie scattering in multi-angle, multi-scale scattering simulations, providing a solid theoretical foundation for high-precision optical measurement and imaging applications.

## 6. Conclusion

We established a GPU-accelerated Monte Carlo-Rigorous Mie scattering transport model for complex environments. Through CUDA parallel architecture optimization, it achieved processing capability of 4.56 million photons per second, realizing high-precision modeling and efficient simulation of light scattering processes in turbid media, providing a solid technical foundation for large-scale accurate simulation.

Through direct quantitative comparison with experimental data, the rigorous Mie scattering phase function and H-G phase function simulation data, the systematic advantages of rigorous Mie scattering phase functions in simulation accuracy were demonstrated. The Rigorous Mie scattering model achieved an average 3.62% root mean square error improvement and 6.33% correlation coefficient improvement compared to the H-G phase function. Under different optical depths, Rigorous Mie scattering exhibited lower errors and higher spatial correlation in radial analysis of light intensity distribution. This indicates that Rigorous Mie scattering can more precisely capture fine angular scattering distributions of photons in complex media, particularly under large particle conditions, while H-G approximation has limitations due to its inherent simplification.

This model accurately quantifies the scattering and attenuation mechanisms of turbid media on radiative transfer, providing a solid physical foundation for correcting optical measurement errors caused by medium interference. The complete technical system formed in we, from theoretical modeling through numerical simulation to experimental validation, effectively solves the key problem of optical system accuracy degradation in complex scattering environments. Future work can explore extending this model to different types of scattering particles, more complex medium geometric structures, and multi-spectral, polarized radiative transfer simulations, further expanding the model's applicability and precision levels.

**Funding**

This work was supported in part by the National Natural Science Foundation of China (Grant Nos. 62105305 and 62403440), the Fundamental Research Program of Shanxi Province (Grant No. 20210302123068), the Science and Technology Cooperation Project of Shanxi Province (Grant No. 202404041101020), the Scientific and Technological Innovation Programs of Higher Education Institutions in Shanxi (Grant No. 2022L004), and the Aeronautical Science Foundation of China (Grant No. 202300340U0002).

**Declaration of generative AI and AI-assisted technologies in the writing process**

During the preparation of this work the authors used Google Gemini in order to improve language and readability. After using this tool, the authors reviewed and edited the content as needed and take full responsibility for the content of the publication.